\documentstyle[icproc%
]{article}

\begin{document}


\newcommand{\ra}{\rightarrow}
\newcommand{\ko}{K^0}
\newcommand{\be}{\begin{equation}}
\newcommand{\ee}{\end{equation}}
\newcommand{\bea}{\begin{eqnarray}}
\newcommand{\eea}{\end{eqnarray}}



\def\eq{\begin{equation}}
\def\eqe{\end{equation}}

\def\ba{\begin{array}}
\def\ea{\end{array}}
\def\unit{1 \hskip-.3em \raise2pt\hbox{$ \scriptstyle |$ } }

\let\und=\b 
\let\ced=\c 
\let\du=\d 
\let\um=\H 
\let\sll=\l 
\let\Sll=\L 
\let\slo=\o 
\let\Slo=\O 
\let\tie=\t 
\let\br=\u 


\def\a{\alpha}
\def\b{\beta}
\def\c{\gamma} 
\def\d{\delta}
\def\e{\epsilon}           
\def\f{\phi}               
\def\g{\gamma}
\def\h{\eta}   
\def\i{\iota}
\def\j{\psi}
\def\k{\kappa}                    
\def\l{\lambda}
\def\m{\mu}
\def\n{\nu}
\def\o{\omega}  \def\w{\omega}
\def\p{\pi}                
\def\q{\theta}  \def\th{\theta}                  
\def\r{\rho}                                     
\def\s{\sigma}                                   
\def\t{\tau}
\def\u{\upsilon}
\def\x{\xi}
\def\z{\zeta}
\def\D{\Delta}
\def\F{\Phi}
\def\G{\Gamma}
\def\J{\Psi}
\def\L{\Lambda}
\def\O{\Omega}  \def\W{\Omega}
\def\P{\Pi}
\def\Q{\Theta}
\def\S{\Sigma}
\def\U{\Upsilon}
\def\X{\Xi}
\def\del{\partial}


\def\ca{{\cal A}}
\def\cb{{\cal B}}
\def\cc{{\cal C}}
\def\cd{{\cal D}}
\def\ce{{\cal E}}
\def\cf{{\cal F}}
\def\cg{{\cal G}}
\def\ch{{\cal H}}
\def\ci{{\cal I}}
\def\cj{{\cal J}}
\def\ck{{\cal K}}
\def\cl{{\cal L}}
\def\cm{{\cal M}}
\def\cn{{\cal N}}
\def\co{{\cal O}}
\def\cp{{\cal P}}
\def\cq{{\cal Q}}
\def\car{{\cal R}}
\def\cs{{\cal S}}
\def\ct{{\cal T}}
\def\cu{{\cal U}}
\def\cv{{\cal V}}
\def\cw{{\cal W}}
\def\cx{{\cal X}}
\def\cy{{\cal Y}}
\def\cz{{\cal Z}}


\def\half{{1 \over 2}}

\def\Bf#1{\mbox{\boldmath $#1$}}       
\def\Sf#1{\hbox{\sf #1}}               
\def\TT#1{\hbox{#1}}                   


\def\bop#1{\setbox0=\hbox{$#1M$}\mkern1.5mu
	\vbox{\hrule height0pt depth.04\ht0
	\hbox{\vrule width.04\ht0 height.9\ht0 \kern.9\ht0
	\vrule width.04\ht0}\hrule height.04\ht0}\mkern1.5mu}
\def\Box{{\mathpalette\bop{}}}                        
\def\pa{\partial}                              
\def\de{\nabla}                                       
\def\dell{\bigtriangledown} 
\def\su{\sum}                                         
\def\pr{\prod}                                        
\def\iff{\leftrightarrow}                      
\def\conj{{\hbox{\large *}}} 
\def\lconj{{\hbox{\footnotesize *}}}          
\def\dg{\sp\dagger} 
\def\ddg{\sp\ddagger} 

\def\>{\rangle} 

\def\<{\langle} 
\def\Dsl{D \hskip-.6em \raise1pt\hbox{$ / $ } }



\def\sp#1{{}^{#1}}                             
\def\sb#1{{}_{#1}}                             
\def\oldsl#1{\rlap/#1}                 
\def\sl#1{\rlap{\hbox{$\mskip 1 mu /$}}#1}
\def\Sl#1{\rlap{\hbox{$\mskip 3 mu /$}}#1}     
\def\SL#1{\rlap{\hbox{$\mskip 4.5 mu /$}}#1}   
\def\Tilde#1{\widetilde{#1}}                   
\def\Hat#1{\widehat{#1}}                       
\def\Bar#1{\overline{#1}}                      
\def\bra#1{\Big\langle #1\Big|}                       
\def\ket#1{\Big| #1\Big\rangle}                       
\def\VEV#1{\Big\langle #1\Big\rangle}                 
\def\abs#1{\Big| #1\Big|}                      
\def\sbra#1{\left\langle #1\right|}            
\def\sket#1{\left| #1\right\rangle}            
\def\sVEV#1{\left\langle #1\right\rangle}      
\def\sabs#1{\left| #1\right|}                  
\def\leftrightarrowfill{$\mathsurround=0pt \mathord\leftarrow \mkern-6mu
       \cleaders\hbox{$\mkern-2mu \mathord- \mkern-2mu$}\hfill
       \mkern-6mu \mathord\rightarrow$}
\def\dvec#1{\vbox{\ialign{##\crcr
       \leftrightarrowfill\crcr\noalign{\kern-1pt\nointerlineskip}
       $\hfil\displaystyle{#1}\hfil$\crcr}}}          
\def\hook#1{{\vrule height#1pt width0.4pt depth0pt}}
\def\leftrighthookfill#1{$\mathsurround=0pt \mathord\hook#1
       \hrulefill\mathord\hook#1$}
\def\underhook#1{\vtop{\ialign{##\crcr                 
       $\hfil\displaystyle{#1}\hfil$\crcr
       \noalign{\kern-1pt\nointerlineskip\vskip2pt}
       \leftrighthookfill5\crcr}}}
\def\smallunderhook#1{\vtop{\ialign{##\crcr      
       $\hfil\scriptstyle{#1}\hfil$\crcr
       \noalign{\kern-1pt\nointerlineskip\vskip2pt}
       \leftrighthookfill3\crcr}}}
\def\der#1{{\pa \over \pa {#1}}}               
\def\fder#1{{\d \over \d {#1}}} 


\def\ha{\frac12}                               
\def\sfrac#1#2{{\vphantom1\smash{\lower.5ex\hbox{\small$#1$}}\over
       \vphantom1\smash{\raise.4ex\hbox{\small$#2$}}}} 
\def\bfrac#1#2{{\vphantom1\smash{\lower.5ex\hbox{$#1$}}\over
       \vphantom1\smash{\raise.3ex\hbox{$#2$}}}}      
\def\afrac#1#2{{\vphantom1\smash{\lower.5ex\hbox{$#1$}}\over#2}}  
\def\dder#1#2{{\pa #1\over\pa #2}}        
\def\secder#1#2#3{{\pa\sp 2 #1\over\pa #2 \pa #3}}          
\def\fdder#1#2{{\d #1\over\d #2}}         
\def\on#1#2{{\buildrel{\mkern2.5mu#1\mkern-2.5mu}\over{#2}}}
\def\On#1#2{\mathop{\null#2}\limits^{\mkern2.5mu#1\mkern-2.5mu}}
\def\under#1#2{\mathop{\null#2}\limits_{#1}}          
\def\bvec#1{\on\leftarrow{#1}}                 
\def\oover#1{\on\circ{#1}}                            
\def\dt#1{\on{\hbox{\LARGE .}}{#1}}                   
\def\dtt#1{\on\bullet{#1}}                      
\def\ddt#1{\on{\hbox{\LARGE .\kern-2pt.}}#1}             
\def\tdt#1{\on{\hbox{\LARGE .\kern-2pt.\kern-2pt.}}#1}   


\def\boxes#1{
       \newcount\num
       \num=1
       \newdimen\downsy
       \downsy=-1.5ex
       \mskip-2.8mu
       \bo
       \loop
       \ifnum\num<#1
       \llap{\raise\num\downsy\hbox{$\bo$}}
       \advance\num by1
       \repeat}
\def\boxup#1#2{\newcount\numup
       \numup=#1
       \advance\numup by-1
       \newdimen\upsy
       \upsy=.75ex
       \mskip2.8mu
       \raise\numup\upsy\hbox{$#2$}}


\newskip\humongous \humongous=0pt plus 1000pt minus 1000pt
\def\caja{\mathsurround=0pt}
\def\eqalign#1{\,\vcenter{\openup2\jot \caja
       \ialign{\strut \hfil$\displaystyle{##}$&$
       \displaystyle{{}##}$\hfil\crcr#1\crcr}}\,}
\newif\ifdtup
\def\panorama{\global\dtuptrue \openup2\jot \caja
       \everycr{\noalign{\ifdtup \global\dtupfalse
       \vskip-\lineskiplimit \vskip\normallineskiplimit
       \else \penalty\interdisplaylinepenalty \fi}}}
\def\li#1{\panorama \tabskip=\humongous                
       \halign to\displaywidth{\hfil$\displaystyle{##}$
       \tabskip=0pt&$\displaystyle{{}##}$\hfil
       \tabskip=\humongous&\llap{$##$}\tabskip=0pt
       \crcr#1\crcr}}
\def\eqalignnotwo#1{\panorama \tabskip=\humongous
       \halign to\displaywidth{\hfil$\displaystyle{##}$
       \tabskip=0pt&$\displaystyle{{}##}$
       \tabskip=0pt&$\displaystyle{{}##}$\hfil
       \tabskip=\humongous&\llap{$##$}\tabskip=0pt
       \crcr#1\crcr}}


\def\phil{@{\extracolsep{\fill}}}
\def\unphil{@{\extracolsep{\tabcolsep}}}


\def\NP{Nucl. Phys. B}
\def\PL{Phys. Lett. }
\def\PR{Phys. Rev. Lett. }
\def\Ref#1{$\sp{#1)}$}

\def\to{\rightarrow}


\def\tj{\tilde{\j}}
\def\td{\tilde{D}}
\def\tv{\tilde{\varphi}}
\def\rld{\rlap{\,/}D}

\def\dv{\dot{\varphi}}
\def\dj{\dot{\j}}
\def\bv{\bar{\varphi}}
\def\bj{\bar{\j}}
\def\rld{\rlap{\,/}D}
\def\rla{\rlap{\,A}\bigcirc}
\def\1ov4{{1\over 4}}
\def\bc{\bar{\chi}}
\def\dox{\dot{x}}
\def\dc{\dot{\chi}}
\def\trld{\tilde{\rlap{\,/}D}}


\def\vecnab{\vec{\nabla}}
\def\vx{\vec{x}}
\def\vy{\vec{y}}
\def\arrowk{\stackrel{\rightarrow}{k}}
\def\kbar{k\!\!\!^{-}}
\def\karrow{k\!\!\!{\rightarrow}}
\def\arrowl{\stackrel{\rightarrow}{\ell}}


\def\SAmpl{ \langle z,\bar\eta | \exp \left( - { \b \over
\hbar } \hat H \right) | y,\chi \rangle }
\def\SAmplb{ \langle z | \exp \left( - { \b \over
\hbar } \hat H \right) | y \rangle }
\def\tr{{\rm tr}}
\def\Tr{{\rm Tr}}


\newcommand{\crr}{\left(\frac{-ie}{\hbar c} \right)}
\newcommand{\ec}{(\frac{e}{c})}
\newcommand{\hce}{\left( \frac{i \hbar c}{e} \right)}


\newcommand{\ffi}{\varphi}
\newcommand{\db}{\delta_{\rm BRST}}


\def\pa{\partial}
\def\xx{\times}
\def\dd{\bar{\d}}


\def\dda{\dot{\alpha}} 
\def\ddb{\dot{\beta}}
\def\ddc{\dot{\chi}}
\def\ddd{\dot{\delta}}
\def\dde{\dot{\epsilon}}
\def\ddf{\dot{\phi}}
\def\ddg{\dot{\gamma}}
\def\ddh{\dot{\h}}
\def\ddi{\dot{\i}}
\def\ddj{\dot{\j}}
\def\ddk{\dot{\k}}
\def\ddl{\dot{\l}}
\def\ddm{\dot{\m}}
\def\ddn{\dot{\n}}
\def\ddo{\dot{\omega}}
\def\ddp{\dot{\p}}
\def\ddq{\dot{\q}}
\def\ddr{\dot{\r}}
\def\dds{\dot{\s}}
\def\ddt{\dot{\t}}
\def\ddu{\dot{\u}}
\def\ddx{\dot{\x}}
\def\ddz{\dot{\z}}

\def\ta{\tilde{\a}} 
\def\tb{\tilde{\b}}
\def\tc{\tilde{\c}}
\def\td{\tilde{\d}}
\def\te{\tilde{\e}}
\def\tf{\tilde{\f}}
\def\tg{\tilde{\g}}
\def\th{\tilde{\h}}
\def\ti{\tilde{\i}}
\def\tj{\tilde{\j}}
\def\tk{\tilde{\k}}
\def\tl{\tilde{\l}}
\def\tm{\tilde{\m}}
\def\tn{\tilde{\n}}
\def\tp{\tilde{\p}}
\def\tq{\tilde{\q}}
\def\tr{\tilde{\r}}
\def\ts{\tilde{\s}}
\def\tt{\tilde{\t}}
\def\tu{\tilde{\u}}
\def\tx{\tilde{\x}}
\def\tz{\tilde{\z}}

\def\ome{\omega}

\def\pa{\partial}
\def\del{\nabla}
\def\delbar{\bar{\nabla}}

\def\xx{\times}

\def\ab{\bar{a}}
\def\bb{\bar{b}}
\def\cd{\bar{c}}
\def\db{\bar{d}}
\def\eb{\bar{e}}
\def\fb{\bar{f}}  \def\Fb{\bar{\F}}
\def\kb{\bar{k}}  \def\Kb{\bar{K}}
\def\lb{\bar{l}}  \def\Lb{\bar{L}} 
\def\mb{\bar{m}}  \def\Mb{\bar{M}}
\def\nb{\bar{n}}  \def\Tb{\bar{T}} 
\def\zb{\bar{z}}  \def\Gb{\bar{G}} 
\def\wb{\bar{w}}  \def\Jb{\bar{J}}
\def\pb{\bar{p}}

\def\Pbb{\bar{\P}} 
\def\jbb{\bar{\j}}
\def\qbb{\bar{\q}}
\def\Sbb{\bar{\S}}
\def\Pbb{\bar{\P}}
\def\dbb{\bar{\delta}}
\def\kbb{\bar{\kappa}}
\def\ebb{\bar{\epsilon}}

\def\ua{\underline{a}}
\def\ub{\underline{b}}
\def\uc{\underline{c}}
\def\ud{\underline{d}}
\def\ue{\underline{e}}

\def\uua{\underline{\a}}
\def\uub{\underline{\b}}
\def\uuc{\underline{\c}}
\def\uud{\underline{\d}}
\def\uue{\underline{\e}}

\def\hL{\hat{L}} 
\def\hM{\hat{M}}
\def\hj{\hat{\j}} 
\def\hf{\hat{\f}}
\def\hMb{\bar{\hat{M}}}
\def\hfb{\bar{\hat{\f}}}

\def\dif{\partial}
\def\difb{\bar{\partial}} \def\dbar{\bar{\partial}}
\def\pab{\bar{\pa}}
\def\nonu{\nonumber \\{}}
\def\half{{1 \over 2}}


\talk{TARGET SPACE SUPERSYMMETRIC SIGMA MODEL 
TECHNIQUES\thanks{preprint no:{\sc LBNL-39561,
UCB-PTH-96/48, KUL-TF-96/22}. To appear in 
proceedings of workshop on Gauge Theory, Applied Supersymmetry and
Gauge Theory, Imperial College, July 5--10 1996, and in
the e-proceedings of Strings'96, Santa Barbara, July 15--20 1996 , and 
Argonne Duality Institute, June 27--July 12 1996.} }

\author{JAN DE BOER\thanks{e-mail:{\sc deboer@theor3.lbl.gov}.
  Fellow of the Miller Institute for Basic Research in Science;
  supported in part by the Director, Office of Energy Research,
  Office of Basic Energy Services, of the US Department of Energy
 under Contract DE-AC03-76SF00098 and in part by 
 the National Science Foundation under grant PHY95-14797.}}

\address{Department of Physics, University of California at Berkeley\\
         366 LeConte Hall, Berkeley, CA 94720-7300, USA\\
         and\\
         Theoretical Physics Group, Mail Stop 50A-5101\\
         Ernest Orlando Lawrence Berkeley National Laboratory, Berkeley, CA 94720, USA}

\author{KOSTAS SKENDERIS\thanks{e-mail:{\sc kostas.skenderis@fys.kuleuven.ac.be}.
Supported by the European Commission HCM program CHBG-CT94-0734.}}

\address{Instituut voor Theoretische Fysica,
         Katholieke Universiteit Leuven \\
         Celestijnenlaan 200D, B-3001 Leuven,
         Belgium}


\maketitle

\abstracts{We briefly review the covariant formulation of the Green-Schwarz superstring
by Berkovits, and describe
how a detailed tree-level and one-loop analysis of this model leads,
for the first time, to a derivation of  the low-energy effective action of the heterotic superstring
while keeping target-space supersymmetry manifest. The resulting low-energy theory
is old-minimal supergravity coupled to tensor multiplet. The dilaton is part of the compensator
multiplet.}

\section{Introduction.}

Sigma model techniques provide the most effective way of obtaining the 
low energy effective actions of string theories. 
The latter are of fundamental importance because they do not only provide the 
starting point for string-inspired phenomenology but also set up the right 
framework to address theoretical questions such as the occurence of 
anomalies or the existense of duality symmetries. Furthermore, 
target space effective actions may turn out to be a powerful tool 
to understand the dynamics of $M$ (or $F$) theory by yielding 
worldsheet actions for theories of membranes and other extended 
objects~\cite{ma}.

Two dimensional sigma models describe the coupling of string theory to 
background fields that are the dynamical fields
of the corresponding low energy effective field theory. 
Standard arguments show that the conformal invariance of the sigma model 
yields the field equations of the background fields. However, until recently,
only bosonic background fields could  be coupled to sigma models.
The determination of the fermion terms in the effective action were left to
(more cumbersome) $S$-matrix techniques. 
This imposes certain limitations on the sigma model methods since the 
supersymmetrization of bosonic actions is often neither straightforward nor
unique. 
Ideally, 
one would like to have a formulation that is manifestly covariant under 
all symmetries that the target space theory may have. In particular, one would
like to keep manifest the target space supersymmetry if present.
This would be achieved by having a formulation directly in terms of
target superspace. This formulation would then yield the {\it off-shell}
description of the low energy effective theory. An off-shell description
is important for a variety of reasons. It constrains the allowed types
of matter coupling in the low energy effective action, it provides constraints
on the structure of the higher order $\a'$-corrections, and it provides
the natural framework to study the interplay between dualities and 
supersymmetry, supersymmetry breaking, and non-renormalization theorems.
In addition, it would be preferable to have all SUSY auxiliary 
fields present if the target space action is going to be a  
worldsheet action of another theory.

A manifestly target space supersymmetric formulation of superstrings,
known as Green-Schwarz formalism, has been known~\cite{GS} for more than a 
decade. The problem, however, of quantizing this string  
while keeping all target space symmetries manifest remained unsolved
for a long period, thus preventing the development of manifestly 
supersymmetric sigma model techniques.
It was only after the breakthrough due to Berkovits~\cite{ber1}  
that such calculations were made possible at all. 
Berkovits managed to find a formulation that can be used to
covariantly quantize 4d compactifications 
of heterotic and type II strings. 
His model contains $N=2$ world-sheet supersymmetry, 
is related by a redefinition to the standard $N=1$
$RNS$ description and reduces to the standard light cone Green-Schwarz 
formulation upon gauge fixing. Following up, Berkovits and 
Siegel~\cite{bersie} constructed the corresponding sigma models.
Indirects arguments based om symmetry considerations give the general 
form of the low energy effective action. The precise form of the latter, 
however, only follows from an explicit ``beta function'' calculation. 
It is important to perform such a calculation since not only it pinpoints the
precise low energy model which is of interest for phenomenology but also 
it provides very strong evidence that the structure 
encoded in the corresponding 
sigma model is the correct one. A contraversial issue which is resolved 
with the new supersymmetric techniques is to which supermultiplet the dilaton 
belongs.  A new paradigm seems to emerge~\cite{bersie,wa3}
 according to which the dilaton is 
always part of the compensator multiplets.
This is true in the bosonic string (the kinetic term of the dilaton comes with the wrong 
sign in the low energy  effective action), and it has been
verified by us~\cite{us}
to be also true for the heterotic string. 
In the latter case, the dilaton is shown to 
be part of a chiral multiplet and not of a linear multiplet as it was
erroneously assumed in the past~\cite{action}. The point here is that 
one should correctly identify the dilaton. We take the dilaton  
to be by definition the worldsheet field that counts string loops. This means
that it couples to the world-sheet curvature in the sigma model. 
Then one can correctly identify the multiplet 
to which the dilaton belongs
by following the derivation of the low energy 
effective action. In the case of type II strings there are
two compensators, one which is chiral (a vector multiplet) and another 
which is twisted-chiral (a tensor hypermultiplet). The Euler number couples
to the sum of the a vector and a tensor compensator~\cite{bersie}. 
This contradicts 
the standard folklore that the dilaton is sitting solely in 
a hypermultiplet.
We intend to settle this issue in a future publication~\cite{us2} 
by an explicit calculation of the type described in this contribution.

Recently we performed a detailed study of the heterotic sigma model~\cite{us}. 
We explicitely checked the superconformal invariance up to one-loop in
$\a'$ by perturbatively computing the OPE's of the generators 
of the $N=2$ superconformal algebra in the supergravity background.
The tree-level superconformal invariance yielded the complete 
supergravity algebra, and the one-loop superconformal invariance the 
equations of motion of the low energy effective theory. We then integrated
these equation to obtain the low energy effective action. The resulting 
low energy effective theory is old-minimal supergravity coupled to a tensor 
multiplet.

In this contribution we will give some details of the ``conventional'' 
beta function calculation. This calculation is much simpler than the 
diagramatic evaluation of the OPE's. Checking the finiteness
of the sigma model, however, is in general a weaker condition 
that the condition for superconformal invariance.  


\section{The supersymmetric sigma model}

In this section we briefly review the supersymmetric 
sigma model~\cite{bersie,us} 
for the heterotic string compactified to four dimensions. The action reads
\bea \label{act}
S & =&  \frac{1}{\a'} \int d^2 z ( \frac{1}{2} 
\Pi^{\a\dda} \bar{\Pi}_{\a\dda} + 
 d_{\a} \bar{\Pi}^{\a} + d_{\dda} \bar{\Pi}^{\dda} +
\frac{1}{2} \bar{\Pi}^A \bar{\Pi}^B B_{BA} \nonu
&  & \qquad -\frac{\a'}{2} (\dbar \rho + i\dbar(\phi-\bar{\phi}))
 (\dif \rho + i\dif (\phi-\bar{\phi})  + a_z))
\eea
Here, $\Pi^A=\partial z^M E_M{}^A,\bar{\Pi}^A=\dbar z^M E_M{}^A$, 
where $z^M=(x^M,\theta^{\a},\bar{\theta}^{\dda})$ are the coordinates of flat
4d superspace, and $E_M{}^A$ is the supervielbein. The other target
space superfields are the super two-form $B_{BA}$ and the dilaton
fields $\phi$ and $\bar{\phi}$ that are chiral and anti-chiral superfields.
The fields $d_{\a},d_{\dda}$ are anti-commuting world-sheet fields
of conformal weight $(1,0)$, and $\rho$ is a boson subject to the constraint
$\dbar \rho + i\dbar(\phi-\bar{\phi})=0$, which is imposed by the Lagrange
multiplier $a_z$. In principle, we still have to add to (\ref{act}) a piece
containing the compactification dependent states and the Yang-Mills
fields, but we will restrict our attention to the pure 4d superfields only.

One crucial feature of (\ref{act}) is that it is manifestly target-space 
supersymmetric. In addition, the dilaton is part of a separate superfield
and does not belong to the same superfield as the antisymmetric
tensor, in contrast to what is sometimes claimed. The way the dilaton
enters is dictated by the world-sheet symmetries of  (\ref{act}).
In the anti-holomorphic sector, it has a Virasoro algebra, but in
the holomorphic sector it has an $N=2$ algebra. The dilaton enters
by means of a generalization of the Fradkin-Tseytlin term~\cite{ft} to
$N=(2,0)$ world-sheet supergravity. The $N=2$ algebra of the sigma model
can be used to prove its equivalence with the usual RNS formulation
of the heterotic string. The BRST cohomology of the sigma model
in flat space yields the same spectrum as that of the conventional
RNS string, and to first order in the background fields the sigma model is given
by its flat space form plus the corresponding vertex operators. 
This demonstrates that it correctly describes the heterotic string.
The sigma model looks somewhat similar to the Green-Schwarz
superstring, however there are important differences. The sigma
model given above does not have $\kappa$-symmetry, and it
turns out that the background fields are not put on-shell already
at tree-level. Without this property, the sigma model would not
be a proper off-shell description of the heterotic string. In addition,
in (\ref{act}) one finds the chiral boson $\rho+i(\phi-\bar{\phi})$,
which does complicate matters somewhat.

The $N=2$ algebra of (\ref{act}) has generators
\bea 
J & = & -i\dif\rho + \dif \phi - \dif \bar{\phi} \nonu
G & = & \frac{1}{i\a' \sqrt{8\a'}} e^{i\rho} d^{\a} d_{\a}  - \frac{1}{\sqrt{2\a'}i} \dif 
\left( e^{i\rho} d^{\gamma} \del_{\gamma} \phi \right)  \nonu
\bar{G} & = &  \frac{1}{i\a' \sqrt{8\a'}} e^{-i\rho} d^{\dda} d_{\dda}   - \frac{1}{\sqrt{2\a'}i} \dif 
\left( e^{-i\rho} d^{\ddg} \del_{\ddg} \bar{\phi} \right) \nonu
T & = & 
 \frac{1}{\alpha'} \left(
 -\frac{1}{2} \Pi^{\a\dda} \Pi_{\a\dda}  - d_{\a} \Pi^{\a}
- {d}_{\dda} \Pi^{\dda}  \right. \nonu
& & \left. 
+ \frac{\alpha'}{2} \dif (\rho+i(\phi-\bar{\phi}))  
   \dif (\rho+i(\phi-\bar{\phi}))
\right) - \frac{1}{2} \dif^2 (\phi+\bar{\phi}).
\label{generators}
\eea

In the case of the bosonic string, where one only has a Virasoro algebra,
the equations of motion of the low-energy effective action follow
by requiring that the sigma model is conformal. This can be checked
by computing its beta-functions. In our case, we want something 
stronger, namely that (\ref{act}) is $N=2$ superconformal. Since it
is not possible to write (\ref{act}) as an action in $N=(2,0)$ world-sheet
superspace, a conventional beta-function calculation would only
guarantee conformal invariance. In order to check to full $N=2$
superconformal invariance, we used a different method, first
applied to the bosonic string by Banks et al.~\cite{banese}. 
This method consists of  a direct computation of the operator
product expansions of the generators of  the $N=2$ algebra, using
a covariant background field formalism. In the case of the bosonic
string, no information is obtained classically and at one loop one
obtains the equations of motion of the low-energy effective field
theory. In our case, we do obtain a set of constraints classically,
that reduce the field content from the supervielbein $E$ and the
super two-form $B$ to that of conformal
supergravity coupled to a tensor multiplet.
These constraints are similar to the so-called
beta-function favored constraints~\cite{gates}.
Together with the Bianchi identities they completely
determine the supergravity algebra, with the result
\bea
\{\del_{\a}, \del_{\b}\} &=& 0,
\label{bian1} \\
\{\del_{\a}, \del_{\ddb} \}
&=& -2i \del_{\a \ddb} -4i H_{\ddb \g} M_{\a}{}^{\g}
+4i H_{\ddg \a} M_{\ddb}{}^{\ddg} +4i H_{\ddb \a} Y, 
\label{bian2} \\
\left[\del_{\a}, \del_{b} \right] &=&
-2 \del_{\b} H_{\ddb \g} M_{\a}{}^{\g}
\nonumber \\
&\ &+[-2i C_{\a \b} \bar{W}_{\ddb \ddg}{}^{\ddd}
+ C_{\ddb \ddg} (\del_{(\a} H^{\ddd}{}_{\b)} -
\frac{1}{3} C_{\a \b} \del^{\e} H^{\ddd}{}_{\e})] M_{\ddd}{}^{\ddg}
\nonumber \\
&\ &+ 2 \del_{\b} H_{\ddb \a} Y  \label{bian3} \\
\left[\del_{a}, \del_{b} \right] &=&
\{-2 H_{\dda \b}  \del_{\a \ddb} \nonumber \\
&\ &+[\frac{i}{2} C_{\a \b} \del_{(\dda} H_{\ddb)}{}^{\g}
+C_{\dda \ddb}(-\frac{i}{6} C^{\g}{}_{(\a|} \del^{\dde} H_{\dde| \b)}
+ W_{\a \b}{}^{\g})] \del_{\g} \nonumber \\
&\ &+\left[ C_{\dda \ddb} \left(\frac{1}{24} \del_{(\a} W_{\b \g}{}^{\d)}
+\frac{1}{4}
(C^{\d}{}_{\a} \del_{(\b|\dde} H^{\dde}{}_{|\g)} + \a \leftrightarrow \b)
\right. \right.
\nonumber \\ 
&\ &
\left. \left.
+\frac{i}{6} C_{\a \g} C^{\d}{}_{\b} \del^{\dde} \del^{\e} H_{\dde \e}
\right)
+\frac{i}{2} C_{\a \b} \del_{\g} \del_{(\dda} H_{\ddb)}{}^{\d}\right]
M_{\d}{}^{\g} \nonumber \\
&\ &-\frac{i}{2} C_{\a \b} \del^{\d} \del_{(\dda} H_{\ddb)\d} Y + {\rm c.c.} \}
\label{bian4}
\eea
where $Y$ and $M_A{}^B$ are the generators of $U(1)$ and Lorentz
transformations, and
 $W_{\a \b \g}$ is completely symmetric chiral superfield,
\be
\del_{\ddd} W_{\a \b \g} = 0,
\ee
and $H_{\dda \b}$ is
defined as follows in terms of the field strength $H_{ABC}$ of $B_{AB}$
($C_{\a\b}$ is proportional to the Pauli matrix $\sigma_2$ and used
to raise and lower indices),
\be
H_{a b c} = C_{\g \a} C_{\ddg \ddb} H_{\dda \b} -
C_{\g \b} C_{\ddg \dda} H_{\ddb \a} \label{habc}.
\ee
$W_{\a \b \g}$ and $H_{\dda \b}$ satisfy the following differential
relations
\be
\del_a H^a = 0, \
\del^{\g} W_{\g \a \b} = \frac{i}{6} \del_{(\a|} \del^{\ddg} H_{\ddg| \b)}
+\frac{i}{2} \del^{\ddg} \del_{(\a|}  H_{\ddg| \b)}, \
\del^{\b} \del_{\b} H_a = 0.
\ee
Furthermore, all components of the field strength $H_{ABC}$ vanish,
except $H_{\a\ddb c}$ and $H_{abc}$.
A similar supergravity algebra has been obtained in \cite{gates}.

At one-loop we found a set of field equations and
the action that yields them. All equations do in fact
follow by requiring that the OPE of $T$ with itself is
correct through one loop. This implies that we can 
also obtain the field equations by a conventional
beta-function calculation. It is not clear whether this
is also true at higher loop, but if true would dramatically
simplify the analysis of higher order corrections to
the field equations.

\section{Conventional $\beta$-function calculation}

For simplicity, we take only the first term of (\ref{act}) and couple it
to an arbitrary world-sheet metric $g_{ij}$,
\be \label{act2}
S=\frac{1}{8\a'} \int d^2 z \sqrt{g} g^{ij} \dif_i z^M E_{M}{}^a \dif_j z^N E_N{}^a + 
    \frac{1}{8} \int d^2 z \sqrt{g} R (\phi+\bar{\phi})
\ee
Next, we take the background metric purely 
conformal, $g_{ij}=e^{\sigma} \delta_{ij}$,
and write the action in $2-2\epsilon$ dimensions,
\bea \label{act3}
S & = & \frac{1}{8\a'} \int d^{2-2\epsilon} z e^{-\epsilon \sigma} 
\delta^{ij} (\dif_i z^M E_M{}^a \dif_j z^N E_N{}^a  \nonu
& & \qquad \qquad +
\a'(\phi+\bar{\phi}) ((1-2\epsilon) \dif_i \dif_j \sigma 
-\frac{\epsilon}{2} (1-2\epsilon) 
 \dif_i \sigma \dif_j \sigma)).
\eea
Conformal invariance means that the action should be 
independent of $\sigma$. Classically,
when the dilaton terms do not 
contribute, this is obvious by taking $\e \rightarrow 0$. At one-loop,
there are counterterms that have to be added to the action in order to 
make the path
integral finite. Such counterterms are needed to 
cancel UV divergences. The only UV
divergences at one-loop come from terms in the background field expansion of
the action of the form $\int d^2 z y^a y^b C_{ba}$, where $y^a$ 
represents a quantum
fluctuation of $z^M$ and $C_{ba}$ depends only on the background fields. 
Contracting the fields $y^a$ and $y^b$ yields a momentum space integral of
the form $\int d^2 k |k|^{-2}$. In principle, in dimensional 
regularization, this
integral vanishes, as the IR and UV divergences cancel each other. However,
string theory should properly be defined on a compact world-sheet in
which case there should be no IR divergence. Therefore, the IR divergence
should be subtracted out, 
yielding~\cite{bmp} $\int d^2 k (|k|^{-2} + \frac{1}{\epsilon} 
\delta^{(2)}(k))=\frac{1}{\epsilon}$. This shows that we have to add to
$\int d^2 z y^a y^b C_{ba}$ the 
counterterm $-\int d^2 z \frac{\a'}{\epsilon} C_{aa}$.
Inserting this in (\ref{act3}) we find a contribution
\be \label{act4}
S  =   \frac{1}{\a'} \int d^{2-2\epsilon} 
z e^{-\epsilon \sigma} ( y^a y^b C_{ab} - \frac{\a'}{\epsilon} C_{aa}
  +\frac{\a'}{8}(\phi+\bar{\phi}) 
\dif_i \dif_j \sigma +{\cal O}(\epsilon)).
\ee
If we take the derivative of this action with 
respect to sigma and then take $\epsilon\rightarrow 0$,
we obtain a nontrivial identity
\be \label{id1}
C_{aa} + \frac{1}{2} \dif \dbar (\phi+\bar{\phi}) = 0.
\ee
If we extract $C_{aa}$ from the background field expansion of the sigma model, and
manipulate (\ref{id1}) using the equations of motion of the sigma model
and the supergravity algebra (\ref{bian1})-(\ref{bian4}), we find three
equations. These are not all independent, but are all consequences of
just one which reads
\be 2 \nabla_{\a} H_{\dda\beta} = \nabla_{\beta} \nabla_a (\phi+\bar{\phi}) .
\ee
This equation of motion can be derived from the action
\be \label{action}
S=\int d^4 x d^4 \theta E^{-1} e^{\phi+\bar{\phi}}
\ee
which is the low-energy effective action of supergravity coupled to a tensor multiplet $G$,
in the gauge where $G=1$. One can verify that the kinetic term of  $\phi$ and $\bar{\phi}$ 
is negative in (\ref{action}), showing that the dilaton is part of the compensator. Thus,
the compensator consists of a chiral and anti-chiral multiplet and we are dealing with
old-minimal supergravity. 

\section{Conclusions}

We have seen how the tree-level and one-loop analysis determine the
off-shell formulation of supergravity corresponding to the heterotic
string. The anti-symmetric tensor is part of a tensor multiplet
with reduced field strength $G$, and the sigma model automatically
selects the 'string gauge' $G=1$. One could reintroduce the
tensor multiplet, leading to the conformally invariant action 
\be S=\int d^4x d^4 \theta E^{-1} G^{-\frac{1}{2}} e^{\phi+\bar{\phi}}. \ee
Thus, the most natural framework for string phenomology 
is one involving matter in a tensor multiplet 
(properties of such actions have been 
 studied in the literature~\cite{action,phenom}), with a chiral
and anti-chiral compensator. However, one should be careful 
not to identify the lowest component of this tensor multiplet
with the dilaton.

Besides this, it would be interesting to include the 
Yang-Mills and compactification dependent fields in the
discussion, and to analyze the type II string.
In addition, we would like to understand whether the
standard beta-function calculation is always
sufficient, and whether it is possible to derive
the low-energy effective action directly from
the sigma model as can be done for the bosonic
string~\cite{tsey}. 

We also hope to apply these supersymmetric 
sigma models to other cases, like the $(1,2)$
heterotic strings~\cite{ma}, and have indications
that it may be possible to find a sigma model
describing self-dual supergravity~\cite{us2}.  


\section*{References}
\newcommand{\Journal}[4]{{#1} {\bf #2}, #3 (#4)}
\newcommand{\NCA}{\em Nuovo Cimento}
\newcommand{\NIM}{\em Nucl. Instrum. Methods}
\newcommand{\NIMA}{{\em Nucl. Instrum. Methods} A}
\newcommand{\NPB}{{\em Nucl. Phys.} B}
\newcommand{\PLB}{{\em Phys. Lett.}  B}
\newcommand{\IJMP}{{\em Int. J. Mod. Phys.} A}
\newcommand{\PRL}{\em Phys. Rev. Lett.}
\newcommand{\PRD}{{\em Phys. Rev.} D}
\newcommand{\ZPC}{{\em Z. Phys.} C}
\newcommand{\SJNP}{{\em Sov. J. Nucl. Phys.}}

\end{document}